\documentstyle[aps,multicol,epsf]{revtex} 
 \def\eqbegin {
\begin{eqnarray} } \def\eqend { \end{eqnarray} }
\def\beq{\begin{equation}} \def\eeq{\end{equation}}
\def\upa{\uparrow}
\def\downa{\downarrow}
 \def\lamda { \lambda } \def\del
{ \partial }

\def\hs_2{\hspace{2mm}}
\def\hs_3{\hspace{3mm}}

\title{Supersymmetry and $d$-Wave Superconductivity}

\begin{document}
\draft 
\maketitle
\begin{center}
{ Kazusumi Ino} \\
{\em Nomura Research Institute,Hongo 2-2-9,Bunkyo-ku,Tokyo,113-0033,Japan}
\end{center} 
\begin{abstract}
Motivated by a recent development in the field theory of 
 the fractional quantum Hall effect, 
we propose a supersymmetric field theoretical model 
of quantum critical $d$-wave and  $d+id$-wave superconductors.  
New concept is a composite particle with the supercharge   
which is formed by electron (hole) and  supersymmetric collective 
 configurations of spin and charge. 
Quantum critical  $d$-wave superconductor is 
characterized as the condensate of these composite particles.   
\end{abstract}

\pagenumbering{arabic}
\begin{multicols}{2}
{\it Introduction. }
High $T_c$ superconductor \cite{bm} 
is a strongly correlated system 
 exhibiting strange behaviors which cannot be 
captured by the Fermi liquid theory where 
 quasiparticles are screened 
enough to be weakly coupled.   
A key basis of strong correlation 
is  low-dimensionality :  the transport property is 
dominated by the ${\rm CuO}_2$ planes between 
insulating layers. The most crucial consequence of 
low dimensionality is the strong Coulomb interaction.    
In two dimensions, the Coulomb interaction remains 
 much stronger than in three dimensions
because the constraint to the plane prevents the 
electrons (holes) to be  screened enough.     
This is actually the essential cause of the fractional 
quantum Hall effect (FQHE). There, instead of screening, 
 electrons  non-perturbatively capture a vortex, 
 forming composite fermions.    
This mechanism enables electrons to 
recoil each other by acquiring 
non-zero relative angular momentum,  
whereby the Coulomb energy is lowered.    
Although FQHE takes place in a strong magnetic field, 
this logic  is by no means  restricted to such a situation.

Besides the Coulomb interaction, antiferromagnetic (AF) 
interaction is believed to be important in 
high $T_c$ superconductor. 
 The parent compound of 
high $T_c$ superconductor is an insulator with the long-range 
antiferromagnetic order. 
Only a small amount of doping  is necessary to destroy
the order. 
Doping introduces holes on the ${\rm CuO}_2$ planes which 
become the charge carrier. From the logic above, 
these holes should capture a vortex to lower the Coulomb energy. 
However, as these holes also have spin, it seems sensible to 
consider that the captured vortex also has the flux of spin (spin flux).
Such a flux is carried by spin-texture. 
This kind of topological doping idea has been considered in   
 some studies \cite{wiegmann,marino,berciu,bfn,senthil}. 
It naturally explains the fact that 
a small portion of doing is enough to destroy the AF order since 
such a topological charge can flip the spins globally. 
Also, in the AF background, to avoid to flip many spins,
the configuration with spin flux is expected 
to have a tendency to form a pair or stripe.
In Ref.\cite{bfn}, the capturing mechanism of spin flux 
 is formulated as 
 an abelian Chern-Simons (CS) gauge theory interacting with 
 a complex boson which is akin  to the field theory of FQHE. 
In this model, superconducting phase is realized as the paired 
condensate of composite fermionic particles with the spin flux. 
The pairing symmetry is $p+ip$-wave. 
On the FQHE side, this model corresponds 
to the Halperin 331 state  \cite{halp}.

Various experiments on high $T_c$ superconductor confirm 
that the actual pairing symmetry  is 
$d$-wave. This pairing symmetry leads to gapless quasiparticles 
residing at the nodes of the pair wavefunction.
Recent photoemission \cite{valla} and Thz conductivity experiments 
\cite{corson}    
have indicated an anomalously short lifetime for the fermionic quasiparticles 
near the nodes. 
A natural explanation below $T_c$ is due to a proximity to a 
quantum phase transition \cite{sachdev} to  some other state \cite{vojta}. 
The global symmetry and field theoretical considerations \cite{vojta} 
restrict the state to be the $d+id$ state \cite{laughlin}. 
Under the existence of such a quantum phase transition, 
the finite temperature property of high $T_c$ superconductor 
may be dictated by the quantum critical point.  
In particular, high critical temperature may be sustained 
 by the quantum critical point.

It is tempting to ask whether it is possible to synthesize 
the topological idea with the quantum phase transition. 
Because the $d+id$ state is $T$-violating, the spin flux attachment 
via Chern-Simons gauge theory may be playing a  role. 
However $d$-wave pairing symmetry is outside the reach of 
abelian Chern-Simons theory. On the FQHE side, 
an example of  $d$-wave paired FQH state is 
known as the Haldane-Rezayi (HR) state \cite{halrez} 
which was originally proposed to solve the $\nu=5/2$ puzzle \cite{clark}.
The HR state is  quantum critical \cite{readgreen} with 
 a neutral massless excitation  of spin 1/2.  
Recently a supersymmetric quantum field theory 
of the HR state has been formulated \cite{ino}. 
It is a certain 2+1 dimensional super Chern-Simons 
gauge theory. In this theory, the topological charge dual to spin  is 
not just a  spin flux, but a super flux with an additional 
 fermionic flux associated with the supercharges.  
These fluxes lead to the quantum critical 
$d$-wave pairing of composite fermions. 
Quantum criticality is tied to  
 the local supersymmetry breaking in this  perspective.

In this paper, we  extend these concepts to  a model of 
high $T_c$ superconductor by 
extending the spin flux attachment to the super flux attachment. 
Namely we consider a picture that holes introduced in the ${\rm CuO}_2$ 
plane by doping will capture a supersymmetric vortex 
with the super  flux.  
Somewhat puzzling appearance of two abelian 
gauge symmetries for  the HR state noted in \cite{ino}
 turns out to have a natural physical meaning. The $d+id$ state  
will be characterized as 
the condensate of composite particle with the flux associated 
with these gauge symmetries.

{\it The Theory. }
For convenience, we will call the charge carrier as {\it electron} hereafter. 
We first present a preliminary observation on the ground state (GS) 
wave function $\Psi_{\rm GS}(x_1,x_2,\cdots)$ of $2N$ electrons  
 in two dimensions with the Coulomb interaction.
The vortex capturing mechanism described above results 
in the zero point structure of $\Psi_{\rm GS}$. 
The most convenient way to express the structure is through 
 Operator Product Expansion (OPE).  
We assume that $\Psi_{\rm GS}$ has a following OPE, 
\eqbegin 
{\rm lim}_{x_1 \rightarrow x_2} \Psi_{\rm GS}(x_1,x_2,\cdots)
\sim |x_1 - x_2|^{l}\Phi^{12}_{\rm GS}+\cdots  
\eqend 
where $l$ is a nonnegative integer determined only by the 
quantum numbers of particles 1, 2 and 
$\Phi^{12}_{\rm GS}$ is a function 
which does not have a zero point at $x_1 \rightarrow x_2$.  
We also assume that $\Psi_{\rm GS}$ has a factorization into 
the spin and the charge factors
$
\Psi_{\rm GS} = \Psi_{\rm spin}\Psi_{\rm charge} 
$
where $\Psi_{\rm charge}$ is non-singular function while 
$\Psi_{\rm spin}$ can have a polynomial singularity.  
The zeroes of $\Psi_{\rm charge}$ and the singularity of 
$\Psi_{\rm spin}$ are constrained by the condition of 
$\Psi_{\rm GS}$ to stand for the electron system.

With these  assumptions we construct a field theory  where 
 electrons capture a super flux.   For this end, we will directly 
 construct the model with all the required  symmetries rather 
 than take a map of singular gauge transformation. 
Thus, we consider a lagrangian of the form 
${\cal L} = {\cal L}_{\rm spin} +{\cal L}_{\rm charge}.$
${\cal L}_{\rm charge}$ is the lagrangian for 
the charge sector with the Coulomb interaction.  
We consider a theory in which ${\cal L}_{\rm spin}$ has 
the supersymmetry of Ref.\cite{ino}. 
Let us give the supersymmetry algebra. 
It has four generators $P^{\mu},Q,\widetilde{Q},\hspace{2mm}\mu=0,1$, 
satisfying  following commutation relations 
\eqbegin
[P^{\mu},P^{\nu}] =0, \hspace{2mm}
\{Q,Q \} = 2\gamma_{\mu}P^{\mu},\hspace{2mm}
[P^{\mu},Q] =2i\gamma^{\mu }\widetilde{Q},
\label{algebra}
\eqend
where we set
$\gamma^0 =\gamma^1 =-\gamma_0=\gamma_1= 1$. Other commutators 
are all set to  zero. 
These commutation relations satisfy the Jacobi identity and 
form a graded Lie algebra.
We may define an invariant metric by the trace
\eqbegin
{\rm Tr}(P^{\mu}P^{\nu}) =-\frac{1}{2}\eta^{\mu\nu}, \hspace{2mm}
{\rm Tr}(\widetilde{Q}Q) = \frac{i}{2}.
\label{trace}
\eqend
In (\ref{algebra}), supercharges appear unsymmetrically.
To treat them in a symmetric way, we may 
take a basis in spin by
$Q^{\upa}=\frac{1}{\sqrt{2}}(Q+\widetilde{Q}),
\hspace{2mm} Q^{\downa}=\frac{1}{\sqrt{2}}(Q-\widetilde{Q})$.
Similar enlarged gauge symmetry has been  known in
 supergravity \cite{dauria-fre} and also studied in 
 the relation between CS theory and superstring theories
 \cite{green}. Physical meaning of (\ref{algebra}) is that   
$Q$ is the sum of the supercharges for spins, $\widetilde{Q}$ 
is the asymmetry between them, and $P^{\mu}$ expresses 
the effect of the Coulomb interaction on the paired supercharges 
\cite{ino}.

The one-form gauge field has the  expansion
\eqbegin
{\cal A}  = iA^{\mu}P_{\mu} + \psi Q
-\widetilde{\psi}\widetilde{Q},
\eqend
where $A^{\mu}$ is  a U(1) gauge field,   
$\psi$ and $\widetilde{\psi}$ are fermionic gauge fields.
We will couple a non-relativistic   matter 
 to these gauge fields. 
We  consider the minimal supermultiplet 
 which arises from a parametrization of supergroup   
$Z=e^X$ where $X=i\phi_{\mu} P^{\mu}+\frac{1}{\sqrt{2}}\lamda Q
-\frac{1}{\sqrt{2}}\widetilde{\lamda}\widetilde{Q}$. Here  
 $\phi_{\mu}$ is  a real boson, $\lamda$ and $\widetilde{\lamda}$ are 
real fermions.  
The construction based on  real bosons instead of complex bosons 
means that we  deal with a vortex field from the start.  
Locally supersymmetric lagrangian is given as 
nonrelativistic chiral lagrangian: 
\eqbegin
{\cal L}_{\rm spin}&=&2i{\rm Tr}(XZ^{-1}D_tZ) 
-\frac{1}{m}{\rm Tr}(Z^{-1}D^{i}Z Z^{-1}D_{i}Z ) \nonumber \\  
&+& A_{\mu}J^{\mu}_{\rm top}+V(Z)+{\cal L}_{\rm CS}, 
\eqend 
where $D_t=\del_t-{\cal A}_t,D_i=\del_i-{\cal A}_i$.
The first term in ${\cal L}_{\rm spin}$ is fixed by the requirement for 
$\phi_{\mu}$ to have the ordinary non-relativistic kinetic term with mass $m$. 
$J^{\mu l }_{\rm top}=\frac{1}{2}
\epsilon^{lmn}\del_{m}\del_{n}\phi^{\mu}$ 
is the topological charge originated from 
the topology of the target space ( a torus $T^2$). 
$V(Z)$ is an invariant potential for $Z$. 
${\cal L}_{\rm CS}$ is the CS term for ${\cal A}$ 
\eqbegin
{\cal L}_{\rm CS} 
&=& \frac{1}{4\pi}{\rm Tr}\left(  {\cal A}\wedge d{\cal A} + \frac{2}{3} {\cal A}\wedge{\cal A}\wedge{\cal A}\right) \nonumber \\
 &=& \frac{1}{4\pi} \left (\frac{1}{2}A^{\mu} \wedge d A_{\mu}
+ i \widetilde{\psi}\wedge d \psi
+ i\psi\wedge\gamma^{\mu}\psi\wedge A_{\mu} \right ). 
\label{action}
\eqend
The lagrangian ${\cal L}_{\rm spin}$ 
has the local supersymmetry $Z \rightarrow U^{-1}(x)ZU(x)$. 
In components, ${\cal L}_{\rm spin}$ becomes 
\eqbegin 
{\cal L}_{\rm spin} &=&  \phi^{\mu}(i\del_t+\frac{\Delta}{2m})\phi_{\mu} + 
i\widetilde{\lamda}(i\del_t+\frac{\Delta}{2m}) \lamda   
-2\sqrt{2}\lamda\gamma^{\mu}\phi_{\mu}\psi_t \nonumber \\
&-&\frac{1}{m}\{i\del^i\lamda\gamma_{\mu}\lamda A^{\mu}_i 
+\sqrt{2}i(\del^i\lamda\gamma^{\mu}\phi_{\mu}+\lamda\gamma^{\mu}\del^{i}\phi_{\mu})
\psi_i \} \nonumber\\ 
&+&A_{\mu}J^{\mu}_{\rm top}+V(Z) +{\cal L}_{\rm CS}.
\eqend 
In the low-energy limit,  
matter fields and collective excitations 
 are reduced to  the Wilson lines.  
For  $n$ Wilson lines, their current is
a sum of delta functions at their contours
${\cal J} = \sum_{i=1}^{n} j^s T_s \delta(C_i) $
where $T_s$ represents the generators of the gauge group.
Then the physics in the low-energy limit is described by the action  
\eqbegin
{\cal S}_{\rm spin} =  
\frac{1}{4\pi}\int{\rm Tr}\left(  {\cal A}\wedge d{\cal A} + \frac{2}{3} {\cal A}
\wedge{\cal A}\wedge{\cal A}\right)  + \int {\rm Tr}({\cal A}\wedge{\cal J}). 
\label{CS}
\eqend
In this form, the states in this system can be exactly 
studied through  the CS/CFT relation \cite{witten}.  
The Hilbert space of the Chern-Simons theory (\ref{CS})  is 
the space of conformal blocks of $c=-2+2$ conformal field theory on 
the cylinder at the infinity \cite{ino}.  
The $c=-2$ part is realized by the symplectic fermions $\theta^a$, 
while the $c=2$ part is realized by two bosons $\varphi^{\mu}$ which 
together form a complex boson $\varphi^{*}$.   
This CFT has a Parisi-Soulas symmetry $Q^{a}_{\rm g}$ as well as two 
U(1) symmetries  $P^{0}_{\rm g}$ and $P^{1}_{\rm g}$ 
(${\rm g}$ stands for ``global''). 
Configurations  of  the system are classified by their behavior 
at the infinity which is determined by the action of 
$Q^{a}_{\rm g}, P^{0}_{\rm g}$, $P^{1}_{\rm g}$ and U(1) 
symmetry of ${\cal L}_{\rm charge}$. 
We can identify them by the fermions and the charges and denote them as  
$(\psi^{a}_{s},q^0,q^1,q_e)$. 
Here $a$ and $s$ stand for the spin and  the holonomy for the fermion 
respectively. 
We denote $s ={\rm R}$ for the trivial holonomy and 
$s={\rm NS}$ for the holonomy $e^{i\pi}$.

Let us now  describe phases in this system. 
The charge part is implicit but satisfies the conditions described above.

{\it Quantum Critical $d$-Wave. } 
We first consider
the ground state of $2N$ electrons which is obtained as
 a paired collection of $N$ $(\psi^{\upa}_{\rm R},1,-1,1)$ and
$N$ $(\psi^{\downa}_{\rm R},-1,1,1)$. 
The charge part $\Psi_{\rm charge}$ has zero points of degree 
$2$ for this state. 
From the CS/CFT relation of Ref.\cite{ino}, 
$\Psi_{\rm spin}$ can be explicitly  calculated as a conformal block 
of corresponding primary fields. It is given by 
\eqbegin
\Psi_{\rm spin}
&=& \langle \del\theta^{\upa}e^{\varphi^{*}}(z^{\upa}_1)
\del\theta^{\upa}e^{-\varphi^{*}}(z^{\downa}_2)\cdots
\del\theta^{\downa}e^{-\varphi^{*}}(z^{\downa}_{2N})  \rangle  \nonumber \\ 
&=&{\rm det}\left( \frac{1}{(z^{\uparrow}_i-z^{\downarrow}_j)^2}\right).
\eqend
This sate is spin-singlet $d$-wave. The FQH counterpart of this state 
is the HR state \cite{halrez}. 
We can characterize the paired condensate
by gauge symmetry breaking.
In our case, two species of composite fermionic particles condensate.
The  supersymmetry is divided into two parts each of which
vanishes in the configuration of one species of composite particle.
A collection of one species   breaks half
of the supersymmetry.
When both species are  present, both parts of
the supersymmetry are broken.

%$\del \theta^{\uparrow}e^{\phi}e^{i\sqrt{q}\varphi},
%\del \theta^{\downarrow}e^{-\phi}e^{i\sqrt{q}\varphi}$
%i.e.   composite particles in the system.
%the state for

Let us next consider quasiparticles.
For the paired state, the elementary
quasiparticle will appear with magnetic flux halved.
Such quasiparticle also generates a non-trivial
holonomy $e^{i\pi}$ for $A^{0}$ or $\psi^{a}$.  
As shown in Ref.\cite{spin}, the modular invariance of the CFT 
admits the existence of such a configuration
 only for $\psi^{a}$.
Thus the elementary quasiparticle and  the elementary quasihole are
 characterized as the species $(\psi^{a}_{NS},0,0,-\frac{1}{2})$,
$(\psi^{a}_{NS},0,0,\frac{1}{2})$. In the CFT, these  
excitations are created by the twist field $\sigma^a$ 
for $\theta^a$  (with the charge part).

We also note that half of the supersymmetry which
doesn't vanish in the configuration of one species of composite particle
produce a fermionic zero mode in the configuration.
The configuration with  a fermionic zero mode gives 
a  neutral massless quasiparticle with spin $1/2$ 
which is generated by the logarithmic field in the ${\rm CFT}$. 
This is an evidence of quantum criticality of the state
 as argued for the HR  state \cite{readgreen}.

{\it  $d+id$ Wave. } 
 We next consider
the ground state of $2N$ electrons which is obtained as
 a paired collection of $N$ $(0,1,-1,1)$ and
$N$ $(0,-1,1,1)$.  In this state electrons do not 
carry a fermionic charge. Accordingly the fermionic 
gauge fields are decoupled in this state.  
These Wilson lines correspond to 
$\del\varphi^{*}e^{\varphi^{*}}
,\del\bar{\varphi}^{*}e^{-\varphi^{*}}$
in the CFT respectively 
( not  $e^{\varphi^{*}},e^{-\varphi^{*}}$ 
since they have conformal dimension zero). 
This is justified by  the property of these fields that they 
become  the extending fields  which   
gives a consistent $c=2$ rational conformal field theory
 with  the modular invariance.
The charge part $\Psi_{\rm charge}$ has a holomorphic 
 zero points of odd  integer degree  for this state. 
$\Psi_{\rm spin}$ is explicitly given by 
\eqbegin
\Psi_{\rm spin} &=& \langle \del\varphi^{*}e^{\varphi^{*}}(z^{\upa}_1)
\del\bar{\varphi}^{*}e^{-\varphi^{*}}(z^{\downa}_2)\cdots
\del\bar{\varphi}^{*}e^{-\varphi^{*}}(z^{\downa}_{2N})  \rangle  \nonumber \\ 
&=&{\rm per}\left( \frac{1}{(z^{\uparrow}_i-z^{\downarrow}_j)^2}\right).
\eqend
This pairing is spin-triplet $d$-wave 
the form of which is known in the context of FQHE \cite{wenwu}.
In this state, two U(1) symmetries are spontaneously broken. 
Quasiparticles in this system has a half  flux for the combined 
gauge field $\gamma^{\mu}A_{\mu}$. In the corresponding $c=2$ CFT, 
it is created by the twist field for the complex boson $\varphi^{*}$.
This state is gapped without massless neutral excitation. 
Thus it is actually $d+id$-wave.  
For the FQHE side, the emergence of this state in the theory suggests 
the existence of ``spin-triplet hierarchy'' on the HR state \cite{ino2}.

{\it Quantum Critical $p$-Wave.}  
We next consider
the state of $2N$ electrons which is obtained as
 a paired collection of $N$ $(\psi^{\upa}_{\rm R},1,0,1)$ and
$N$ $(\psi^{\downa}_{\rm R},-1,0,1)$.   
The charge part $\Psi_{\rm charge}$ has holomorphic 
zero points of odd integer degree  for  this state. 
The CFT for the spin part  is given by  
 the $c=-1$ $\beta$-$\gamma$ ghost system \cite{FMS} with 
an extension on the zero modes \cite{spin}.    
$\Psi_{\rm spin}$ can be explicitly  calculated as a conformal block 
 of $\beta$-$\gamma$ 
\eqbegin
\Psi_{\rm spin}
&=&\langle \beta(z^{\upa}_1) \gamma(z^{\downa}_2)\cdots
\beta(z^{\upa}_{2N-1})\gamma(z^{\downa}_{2N})  \rangle \nonumber \\
&=& \langle \del\theta^{\upa}e^{\varphi^{0}}(z^{\upa}_1)
\del\theta^{\upa}e^{-\varphi^{0}}(z^{\downa}_2)\cdots
\del\theta^{\downa}e^{-\varphi^{0}}(z^{\downa}_{2N})  \rangle  \nonumber \\ 
&=&{\rm per}\left( \frac{1}{z^{\uparrow}_i-z^{\downarrow}_j}\right).
\label{quantump}
\eqend 
This state is spin-singlet $p$-wave.   
The FQH counterpart of this state is the permanent state
 forming a hierarchy of the HR state \cite{spin,ino}. 
As in the quantum critical $d$-wave state, 
a collection of one species   breaks half
of the supersymmetry.
When both species are  present, both parts of
the supersymmetry are broken.  
Accordingly this state also has neutral massless 
excitations created by logarithmic field in the CFT. 
This is the sign of the quantum criticality as was argued for the quantum 
critical $d$-wave state.

Quasiparticle in this state  generates a non-trivial
holonomy $e^{i\pi}$ for $\psi^{a}$.  
The elementary quasiparticle and  the quasihole are
 characterized as the species $(\psi^{a}_{NS},0,0,-\frac{1}{2})$,
$(\psi^{a}_{NS},0,0,\frac{1}{2})$.  
As in the quantum critical $d$-wave state, these  
excitations are created by the twist field $\sigma^a$ 
for  $\theta^a$  (with a charge part).

The CFT of this state and the CFT for the $d+id$ state have 
the same partition function up to the inclusion of 
 a half  flux \cite{spin}. It suggests a close relation between them.

\begin{figure}
\vspace{0.2cm}
\epsfxsize=3.5in
%\centerline{\psfig{figure=phase.eps,width=7.5cm,angle=0}}
%\epsfile{file=fig1,width=40mm} 
\centerline{\epsfbox{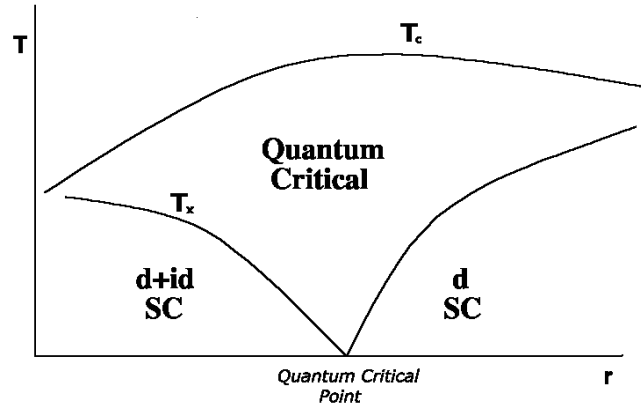}}
\vspace{0.2cm}
\caption{Schematic Phase diagram recently advocated by Vojta et al.}
\label{phases}
\vspace{0.2cm}
\end{figure}

{\it Phase Diagram.}
In Ref.\cite{vojta} the phase diagram 
shown in Fig.\ref{phases} is proposed 
for the $d$-wave superconductor.  It is argued that 
the $d$-wave superconductor is at a proximity to the 
quantum phase transition to the $d+id$ superconductor. 
In Fig.\ref{phases} the parameter $r$ can be 
the doping parameter, the inverse of the strength of the 
external magnetic field and so on.  
    
Let us compare this picture with our theory. 
The $d+id$ superconductor and the quantum critical point 
are identified with  the $d+id$-wave state and 
the  quantum critical $d$-wave state respectively.
Thus neutral massless quasiparticles at 
the 'quantum critical soup' arises from the supersymmetry breaking. 

In our theory, time-reversal symmetry is implicitly broken, 
thus non $T$-violating $d$-wave superconductor is out of 
 reach. One possibility is to consider it as 
the quantum critical $d$-wave state disturbed by the 
underlying square lattice symmetry.  This line of argument has 
been given for the $p$-wave case in Ref.\cite{bfn}.

\begin{figure}
\vspace{0.2cm}
\epsfxsize=3.5in
%\centerline{\psfig{figure=phase.eps,width=7.5cm,angle=0}}
%\epsfile{file=fig1,width=40mm} 
\centerline{\epsfbox{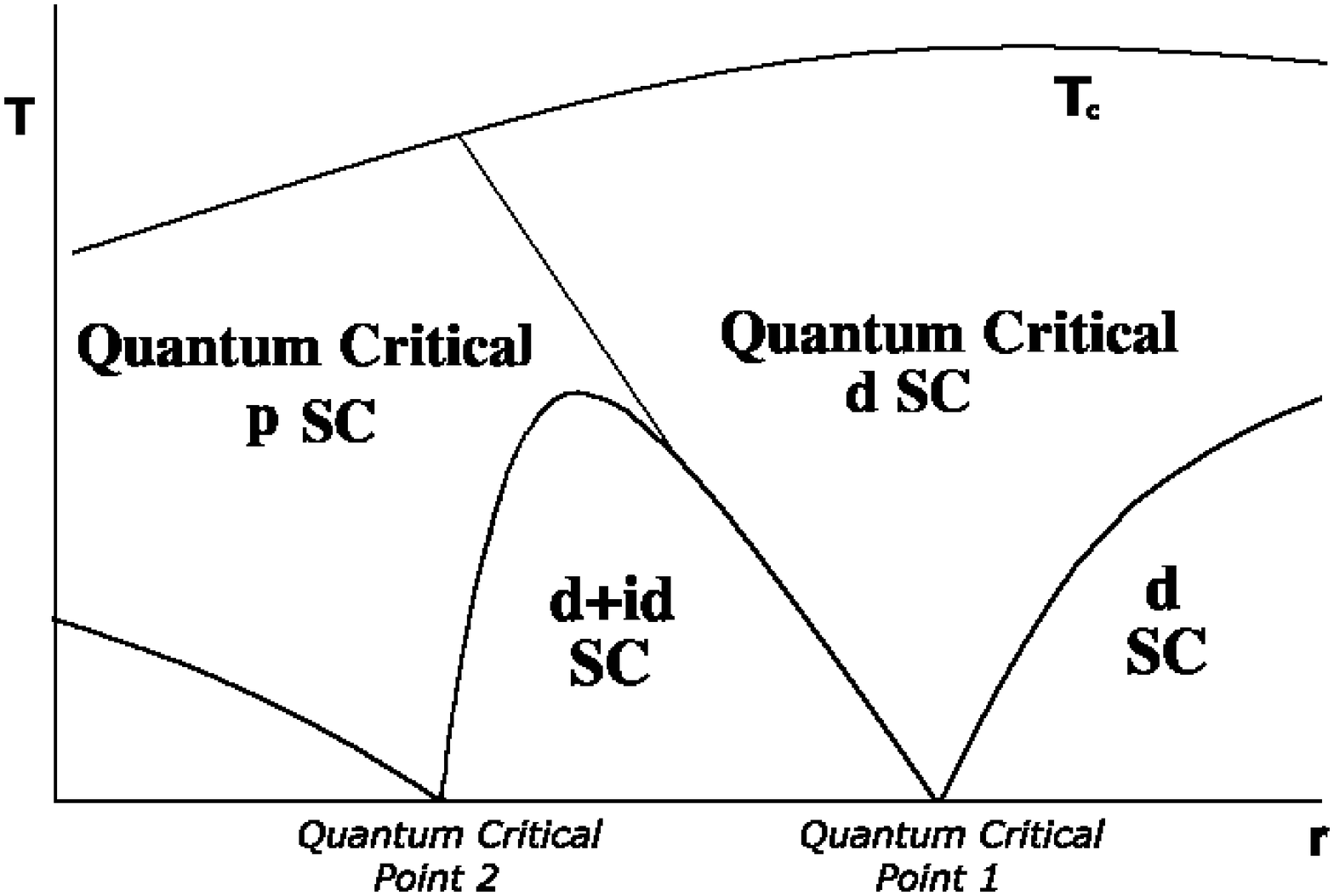}}
\vspace{0.2cm}
\caption{Modified Schematic Phase diagram 1}
\label{phases2}
\vspace{0.2cm}
\end{figure}

Our theory predicts the existence of  
another quantum critical point at which the quantum critical $p$-wave state 
emerges.  
Almost equivalent CFTs for  
the $d+id$ and the quantum critical $p$-wave states suggest that 
the $d+id$ state exposes to the quantum critical point. 
By interpreting  $r$ as the inverse of the strength of the 
 external magnetic field,     
 quantum phase transitions  from $d$-wave $\rightarrow$ 
quantum critical $d$ $\rightarrow$  
$d+id$ $\rightarrow$ quantum critical $p$ 
 is likely to be induced by increasing the  magnetic field strength  
(Fig.\ref{phases2}).  An intriguing aspect of this picture is 
a possible direct transition between two quantum critical states, 
which is compared to  the spin-singlet FQH hierarchy \cite{spin,ino}. 
As noted  in Ref.\cite{ino}, the inverse transition from
 quantum $p$-wave to quantum $d$-wave is 
 achieved  by capturing of the vortex with the unit flux for 
 $A^1$, which is the same as the $p+ip$-wave pairing \cite{bfn}.

\begin{figure}
\vspace{0.2cm}
\epsfxsize=3.5in
%\centerline{\psfig{figure=phase.eps,width=7.5cm,angle=0}}
%\epsfile{file=fig1,width=40mm} 
\centerline{\epsfbox{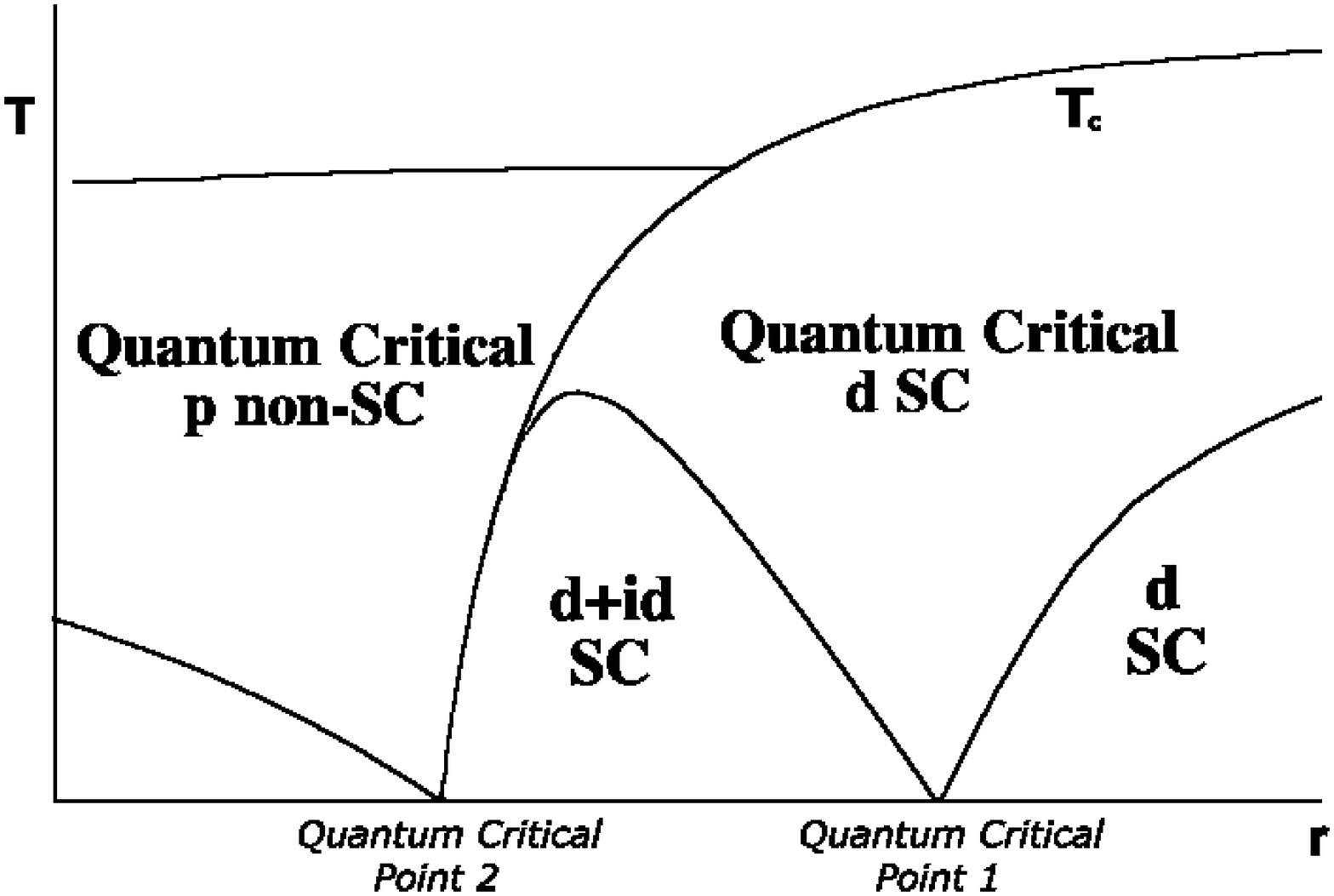}}
\vspace{0.2cm}
\caption{Modified Schematic Phase diagram 2}
\label{phases3}
\vspace{0.2cm}
\end{figure}

Another intriguing aspect  on  the quantum critical $p$-wave state is 
its  relevance to the pseudo-gap state. 
This arises when  we interpret $r$ as  the hole concentration.
To relate the quantum critical $p$-wave state to the psuedo-gap state,
we must consider the quantum critical $p$-wave state without 
the charge condensation (Fig.\ref{phases3}). It means that 
 the condensation occurs only for the spin-part.   
This identification seems natural when one takes into account 
the symmetric staggered  flux in $\Psi_{\rm spin}$ of 
(\ref{quantump})  (the permanent factor) 
 is what is expected for the spin-texture picture 
of doped holes in the AF background.

{\it Discussions.} In this paper, we proposed a supersymmetric field theory 
which naturally embodies  conjectured phases  of 
 the $d$-wave superconductor.    
This model has  rich theoretical contents  rooted in 
 deep  aspects of  topological quantum field theory \cite{witten}, 
supersymmetry \cite{dauria-fre,green} and 
conformal field theory \cite{FMS}.
Although we concentrate on  condensed phases, it is 
 also possible to consider the normal state, providing a realization of 
the spin-charge separation \cite{PWA} in a new way.

The most important question we didn't address in this paper is 
the high critical temperature $T_c$. 
We'd like to  consider it as a property of the field theory 
at the quantum critical point. 
In our model, the key property
 is the local supersymmetry of (\ref{CS}) and the emergence 
of fermionic gauge field  associated with it.  
By the strong suppression through the fermionic gauge field, 
$\Psi_{\rm GS}$ at the 
quantum critical $d$-wave state 
 has no zero points between the paired electrons.  
The same mechanism should  modify the temperature dependence of 
the effective potential of the charges. 
Thus we suggest a  study on 
the beta function of the potential for future work.

{\it Acknowledgement.} The author is  grateful to
 Masaki Oshikawa for helpful discussions.

\vskip 0.2in
\noindent

\def\NP{{Nucl. Phys.\ }}
\def\PRL{{Phys. Rev. Lett.\ }}
\def\PL{{Phys. Lett.\ }}
\def\PR{{Phys. Rev.\ }}
\def\IJMP{{Int. J. Mod. Phys.\ }}

\end{multicols}

\end{document}